\documentclass[fleqn,12pt,letter]{wlscirep}
\title{MWTP: Monterrey Weather, Traffic and Pollution Database for Geospatial Analysis}

\author[1,2]{J. Montalvo-Urquizo}
\author[1,2,*]{M. G. Villarreal-Marroqu\'in}
\author[1,2]{J. J. Hern\'andez-Castillo}
\author[1]{H. E. Hern\'andez-Gonz\'alez}
\affil[1]{Centro de Investigaci\'on en Matem\'aticas, A.C., CIMAT-Monterrey, Mexico}
\affil[2]{C\'atedras CONACYT, Mexico}

\affil[*]{Corresponding author: maria.villarreal@cimat.mx}

\begin{abstract}
The MWTP: Monterrey Weather, Traffic and Pollution Database contains a collection of historic weather conditions, and recent traffic and air pollution data of the metropolitan area of Monterrey, Mexico. Collected data includes: temperature, humidity, raining conditions, wind speed, travel distance, travel time, PM2.5, PM10, O$_3$, among many others. As of February 2017, the MWTP database contained more than 4 million records of weather data, more than 700 thousand records of traffic data and around 57 thousand of air pollution measurements. Here, it is described how the data is been collected and structured into a databased, so it can be used for statistical and geospatial analysis of the Monterrey area.
\end{abstract}

\begin{document}

\flushbottom
\maketitle

\thispagestyle{empty}

\section{Introduction}\label{sec:intro}
Here it is presented the recently launched system for constructing the Monterrey Weather, Traffic and Pollution (MWTP) database, a  comprehensive collection of historical and current data of weather conditions, traffic and air pollution of the metropolitan area of Monterrey, Mexico. The data covers the city of Monterrey which is the state capital of Nuevo Leon (NL), one of the most industrially developed states in Mexico. The Metropolitan Area of Monterrey or AMM (for its Spanish abbreviation) is a conglomerate of 12 municipalities that concentrate 88.5\% \cite{enc2015nl} of the total state population (5.1 millions as of 2015 \cite{enc2015nl}). The AMM also represents the 3\textsuperscript{rd} largest metropolitan area in the country. 

On the 2015 intercensal survey \cite{enc2015nl} it was estimated that  53\% of the population 12 years old and older are economically active, about 2.2 million people. And from the non economically active 32\% \cite{panorama2015nl} are students (about 800k). 65\% of childes age 3 to 5 and 99.7\% of age 6 to 14 attend school, as well as 44\% of the population age 15 to 24. Therefore, close to 3 million people need to commute to work and/or study daily. Thus estimates of school and labor mobility are important for big cities and were included on the 2015 intercensal survey. 

The main modes of transportation used in the AMM to get to work and school are walking, privet car (automobiles, trucks or motorcycles), and public transportation (bus, taxi, metro, metro bus). 45.9\% of the student population 3+ years old walk to school, 28.7\% use privet car, 20.8\% use public transportation, and only 6\% use school transportation. Is is estimated that 58.2\% of these people take up to 15 minutes to go to school; 20.4\% spend 16 to 30 minutes; 11.7\% spend 31 to 60 minutes; and 3.5\% more than an hour. Therefore, most of the people spend up to 30 minutes to go to school. However, if travel time is divided by transportation mode, it is estimated that 87\% of the people that walk to school take up to 16 minutes to get there. And more than 55\% of the students that use privet or public transportation will take form 16 minutes to 1 hour commuting to school. From the population that work and need to go to work (94\%), 38.3\% used private vehicles, 38.6\% used public transportation, 10.4\% walk to work, 8.7\% used personnel transportation, and only 1.8\% used bicycle. 55.8\% of the people take 16 to 60 minutes commuting to work, 19.8\% spend up to 15 minutes, and 12.9\% more than 1 hour. However, more than 65\% of the people that use public, private or personnel transportation commute between 16 and 60. These estimates give a big picture of the daily commute time of the people in the state of NL. 

Commute time is not only impacted by distance if not by traffic congestions. It is estimated that 57\% of the particular houses have a car \cite{enc2015nl}. In 2015 was estimated that NL had 1,865,729 \cite{paginainegi} register vehicles: 1,377,434 are automobiles, 8,374 passenger buses, 427,712 trucks and cargo cars, and 52,209 motorcycles. And the estimated number of particular houses was 1,393,542, from which 86.2\% are on the AMM. 

As can be seen for the previous estimates, most of the people used private car and public transportation to get to work and school, and most of them commute between 16 to 60 minutes. Therefore, air pollution and traffic have become a big issue that requires a delicate analysis to support Public Policy decision making, such as road space rationing similar to the no-drive day program implemented in Mexico City. Whether its the local industry, the automobiles or a combination of these two and other factors who contribute to the raise of air quality indicators remains a subject that needs to be carefully studied in major cities around the word and as is the case in the AMM.

The AMM was recently been reported by the World Health Organization (WHO) as one of the cities with highest pollution levels in Mexico. Therefore, in order to help answerer critical questions about pollution and its possible correlation with traffic levels and weather conditions, on May of 2016 we started to collect public weather, traffic and air pollution data from several station around the AMM. The data is collected in real time and has been structured on a single database to facilitate its analysis. Section \ref{sec:datacollection} of this report details how the data is acquired and Section \ref{sec:database} describes how the database was structured. Last, conclusions and some ideas of possible data uses are described in Section \ref{sec:last}.

\section{Data Collection}\label{sec:datacollection}
In order to make possible the analysis of effects and relations of weather, transportation, and measured pollution, the MWTP database contains a collection of data obtained from different open and/or public access resources. All data share the feature of being geospatially referenced, either through the locations of the weather or pollutants measurements, or by the start and end points of the street routes connecting some points of interest of the AMM. In the forthcoming subsections we present some details of how the data is acquired. 

\begin{figure}[htb]
\centering
\includegraphics[width=\textwidth]{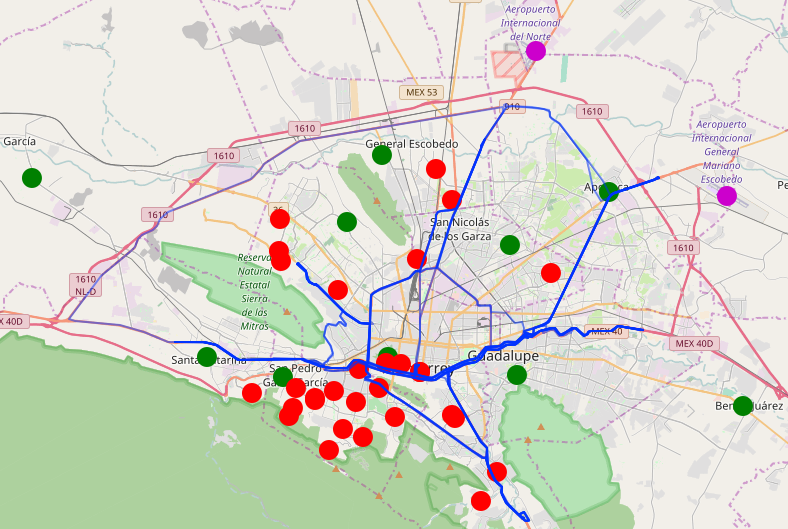}
\caption{\label{fig:mapa} AMM map with PWS (red points), public weather stations (magenta points), air pollution monitoring stations(green points), and vehicle transportation routes (blue lines).}
\end{figure}

\subsection{Weather Data}
Weather data is acquired using the Weather Underground API \cite{wunderwebpage}. A script in R was implemented to download daily historical data using the command \textit{yesterday}. This is, each day historical weather data of the day before is queried. Weather Underground has historical and forecast weather data from more that 200,000 personal weather stations (PWS) around the word. PWS are located on offices, houses, patios etc. and report current weather conditions every 5 to 60 minutes, depending on the station. In addition, it has information of around 12,000 public stations or Meteorological National Stations. The AMM, currently, has 30 PWS and 2 public weather stations (located on two airports). Figure~\ref{fig:mapa} shows on a map the location of the PWS (red) and public weather stations (magenta). Weather conditions such as: temperature, pressure, humidity, precipitation rate, wind direction and speed, among others are collected hourly (or in smaller time intervals) at each station. All the collected weather conditions are shown on Table~\ref{tab:WInfo}.

The MWTP database has historical data since 1973 from the 2 public stations (airports); and for some PWS it has daily records since 2008. Overall, as of February 2017, we have about 4 million records and closet to 40 million data points of weather information. A record of weather is a vector (row of weather table) containing the information of all weather variables (columns or attributes of weather table) on a particular time-day for an specific weather stations.

\subsection{Traffic Data}
As in many other large urban areas, the main streets of the AMM are very busy before and after the common opening hours of most offices and commercial stores. The MWTP collects real time information about the status for connecting several points of the AMM at locations that are important for their practical use like the downtown and the main airport, or because they represent a dense populated area with known traffic jam problems.

The traffic data is collected using the Google Maps API\cite{googlemaps} which allow the real time download of traffic information between two given locations. The calls provide information of typical travel time between the locations, the current estimated travel time based on real time traffic conditions, and the total travel distance (which may vary if a traffic jam is detected and the optimal route differs from the traditional). The traffic information has been continuously collected since May 15\textsuperscript{th}, 2016 and contains around 700 thousand registries until February 2017. A single registry of traffic consists of the time stamp, travel distance, typical travel time, and real-time duration of the trip and number of the corresponding route. The amount of registries is increasing at a rate of 75 thousand new registries per month and the amount of registries will cross the one million mark by July this year.

The routes were selected by choosing seven points inside the AMM and considering all possible connections between them. As some routes are different when taken in the opposite direction, for every selection of two points gives in general two different routes and we have to consider the $7!/(7-2)!=42$ routes. The different distances obtained in the AMM traffic routes vary from 12.3 km to  60.4 km, being the longest distance the one connecting the Mariano Escobedo Airport with the dense populated municipality of Santa Catarina. The blue lines on the map of Figure~\ref{fig:mapa} represent all 42 routes typically taken outside the hours of traffic congestion. During rush hours, the routes may vary due to the possibility of a connecting route with a larger distance but a faster connection.

\begin{figure}[htb]
\centering
\includegraphics[width=.85\textwidth]{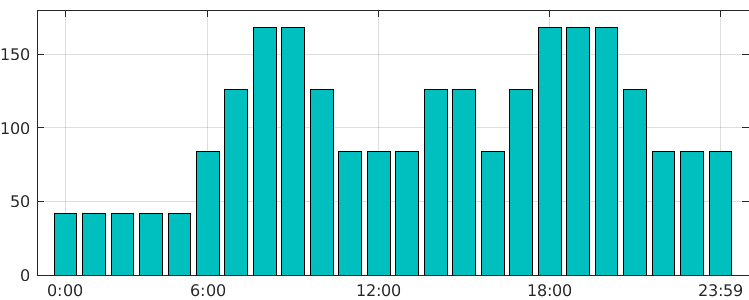} 
\caption{Traffic information data calls during one day.}\label{fig:traffic_timecalls}
\end{figure}

As it is well known that the mornings and evenings are the most problematic times for traffic conditions, the frequencies for collecting the traffic information have bee adjusted to obtain a better time resolution during these time periods. Figure \ref{fig:traffic_timecalls} shows a plot of the hourly calls performed to better catch the effects of peak hours in the AMM traffic conditions.

\subsection{Air Pollution Data}
The Sistema Integral de Monitoreo Ambiental (SIMA) is the public office in the state of NL that monitors and measures air quality in the AMM. It has 10 monitoring stations that can be seen in green on Figure~{\ref{fig:mapa}. Data on air quality is available on real time in \cite{aire}. The stations measure 6 contaminants: PM10 (particulate matter smaller than 10 $\mu$m), Ozone (O$_3$), Carbon Monoxide (CO), Sulfur Dioxide (SO$_2$), Nitrogen Dioxide (NO$_2$). The contaminants are reported using the IMECA (Metropolitan Air Quality Index of Mexico) scale (see \cite{aire} measurement standards) with the following categories:
\begin{itemize}
\item 0-50, air quality is considered GOOD
\item 51-100, air quality is considered REGULAR
\item 101-150, air quality is considered BAD
\item 151-200, air quality is considered VERY BAD
\item 201-500, air quality is considered EXTREMELY BAD
\end{itemize}

In order to get air quality information, a program written in R was used to collect the data directly from the website. By using a direct link, we can access the webpage that contains one table with hourly measurements for one contaminant and one station, from 2:00 am to the hour of the request. The website does not report measurements at 12:00 am and 1:00 am. Then, using the R package XML the tables for every combination station-contaminant are read. After collecting all the readings for a single day, a table is prepared for every station containing time and date, and the readings of all the contaminants. Then ten formatted tables (one for each station) are loaded into the database using an R routine and the MySQL package. This job is scheduled daily.

\subsection{Data Preprocessing}
Before the data is inputed into the database it is preprocessed. Up to this point, we have established upper and lower range for all collected variables based on logical sense. For example, wind direction needs to be between 0 and 360\textsuperscript{o}. If a value out of the range is detected, it is input as 'NA' on the database. However, for some variables has not been possible to indicate a range in advance. Therefore, in the further it will be necessary to implement techniques to detect outlier points before the data is analyzed. 

On the other hand, in some case we have severe missing data, for example, when a pollution station is on maintenance, data is not collected for a long period of  time. For this cases it may be necessary to developed techniques to input missing data.

\section{Database Structure and Content}\label{sec:database}

\subsection{Database structure} 
The MWTP database was implemented as a relational database using MySQL 5.6.32. It runs on a IBM DL380 G9 server with a CENTOS 7 operating system.  R scripts are  used to parse and load data from flat-files (csv) and structure data format (xml) files into the database using a TCP/IP connexion. The database tables are of type InnoDB.

Figure~\ref{fig:entity-rel} shows the entity-relationship MTWP database diagram. The database consists of 10 tables. The 3 major tables are: (1) Weathers, (2) Traffics and (3) Pollutions. The name and description of the attributes (columns) on each table are given in Tables \ref{tab:WInfo}, \ref{tab:TInfo}, and \ref{tab:PInfo} respectively. Each record (set of attributes at a particular time stamp) is related to a location ID that is found on the location\_w, location\_t, or location\_p table for weather, traffic and air pollution location respectively. Location tables include geo-spatial information such as latitude and longitude of weather and air pollution stations and longitude and latitude of star and end point of traffic routes. They also include a short name and description of each location. Table \ref{tab:Locations} shows the attributes of each locational table. This tables help maintain a referenced structure that facilitates search queries, include or delete new station and/or routes.  

Since several of the weather variables are categorical, additional (dependent) tables were used to related the levels of the variables with the main weather table. Table~\ref{tab:ExtraInfW} shows the information of the 4 extra tables (conds, icons, time\_zone and wdires) that are related to weather.

\begin{figure}[h!]
\centering
\includegraphics[width=1\textwidth]{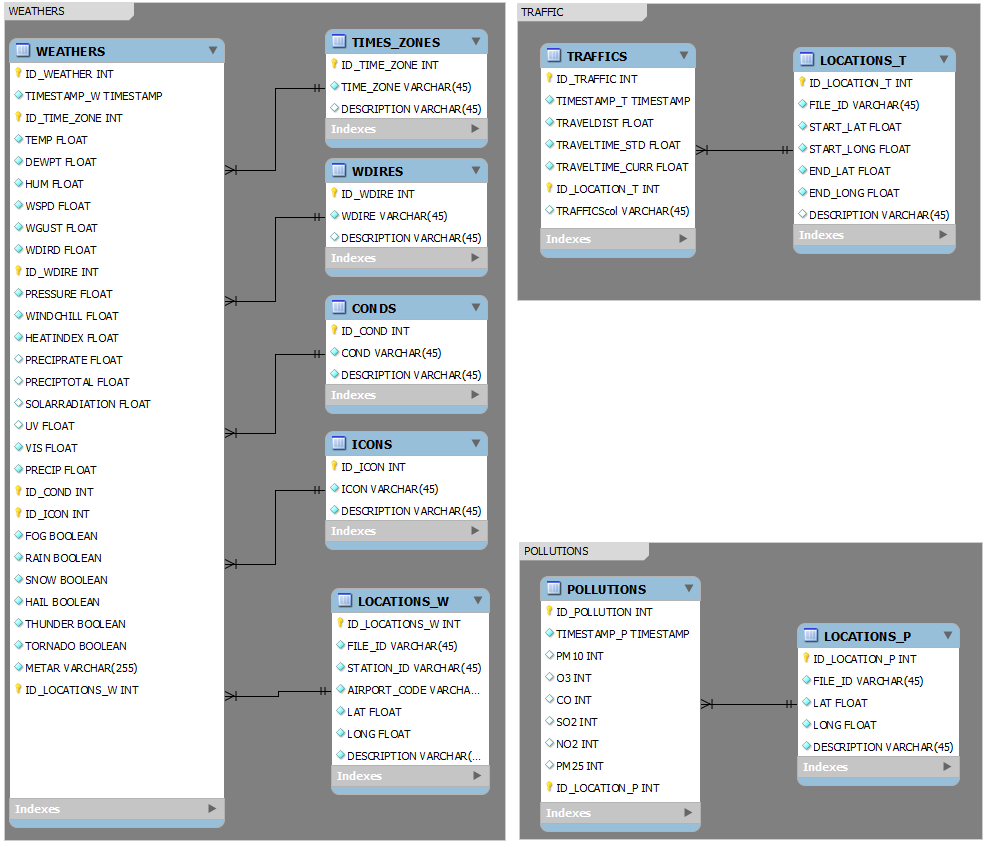}  
\caption{\label{fig:entity-rel} Entity-relationship flow diagram }
\end{figure}

\subsection{Current Content} 
Table~\ref{tab:numrecords} is a summary of the number of non empty records per attribute (as of February 2017). Currently, we have close to 4.1 million records (timestamps-station) of weather data, more than 700K records of traffic (timestamps-routes) and more than 56K records(timestamps-station) of air pollution measurements. The third column of Table \ref{tab:numrecords} shows the average monthly number of records collected per attribute.

\begin{table}[!ht]
\begin{center}
\begin{small}
\begin{tabular}{ l r r}
\hline
Attribute  & Non-empty records & Monthly average\\
\hline
\textbf{weather	}	&	&	\\
~	TEMP		&	4033609	&	94745	\\
~	DEWPT		&	4033181	&	94745	\\
~	HUM		&	4033645	&	94776	\\
~	WSPD		&	2980111	&	57249	\\
~	WGUST		&	1907801	&	39347	\\
~	WDIRD		&	2958867	&	55121	\\
~	ID\_WDIRE		&	4068948	&	94832	\\
~	PRESSURE		&	4065537	&	94561	\\
~	WINDCHILL		&	44054	&	1339	\\
~	HEATINDEX		&	979480	&	1104	\\
~	PRECIPRATE		&	2331574	&	53306	\\
~	PRECIPTOTAL		&	2655735	&	56912	\\
~	SOLARRADIATION		&	667058	&	21152	\\
~	UV		&	605462	&	19011	\\
~	VIS		&	512502	&	1302	\\
~	PRECIP		&	6	&	0	\\
~	FOG		&	555199	&	1302	\\
~	RAIN		&	555199	&	1302	\\
~	SNOW		&	555199	&	1302	\\
~	HAIL		&	555199	&	1302	\\
~	THUNDER		&	555199	&	1302	\\
~	TORNADO		&	555199	&	1302	\\
\textbf{traffic	}	&	& \\
~	TRAVELDIST		&	657704	&	74227	\\
~	TRAVELTIME\_STD		&	657704	&	74227	\\
~	TRAVELTIME\_CURR		&	657704	&	74227	\\
\textbf{pollution}		&	&	\\
~	PM10		&	42291	&	5037	\\
~	O3		&	29695	&	3274	\\
~	CO		&	31391	&	3565	\\
~	SO2		&	38317	&	4391	\\
~	NO2		&	25558	&	2527	\\
~	PM25		&	13018	&	484	\\	
\hline
\end{tabular}
\end{small}
\end{center}
\caption{Summary of non-empty records on MWTP Database up to February 9, 2017, and the average monthly number of records per attribute.}
\label{tab:numrecords}
\end{table}

\clearpage
\subsection{Data Selection} 
Since the MWTP Database is structured, we can select data of a particular variable (attribute) for all stations during a particular interval of time, day-time, or we can get the data of all (or some) attributes of a station (or some) during a period of time, among many others.

As an example, Figure~\ref{fig:temp_2016} shows graphically the data obtained of a search of temperature (TEMP) at the two airport stations (ID\_LOCATION\_W =  31 AND 32) for the year 2016 (TIMESTAMP\_W BETWEEN 2016-01-01 AND 2016-12-31). This search generated 16,088 observations (rows) and 3 variables (temp, timestamp, location\_id) and it took 2.5645 seconds.

\begin{figure}[h!]
\centering
\includegraphics[width=0.95\textwidth]{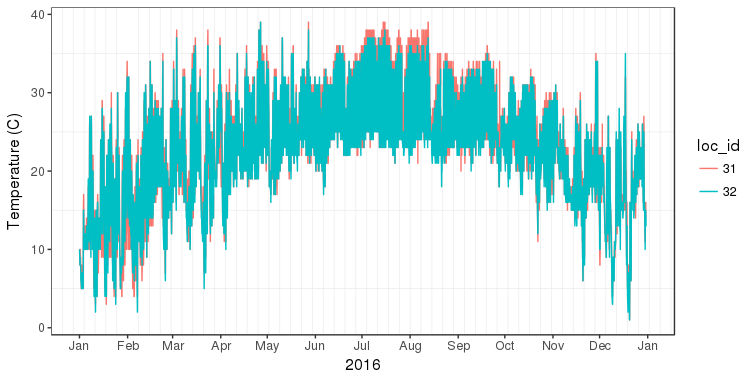}  
\caption{\label{fig:temp_2016} Records of temperature at airport stations during 2016.}
\end{figure}

An example of data extracted from traffic information is shown in Figure \ref{fig:morones-const}. The plots in Figure \ref{fig:morones-const} show the curves for two routes corresponding to the road pair Constituci\'on-Morones Prieto. These  are the main roads connecting east and west of the AMM, with a connecting distance of around 35 km. The data shown in the plots were obtained making a search over the current travel time (TRAVELTIME\_CURR) for the routes (ID\_LOCATIONS\_T =  11 AND 32) for the third week of February 2017 (TIMESTAMP\_W BETWEEN 2017-02-13 AND 2017-02-17). The searched includes 295 observations per route. 

\begin{figure}[h!]
\centering
\includegraphics[width=0.95\textwidth]{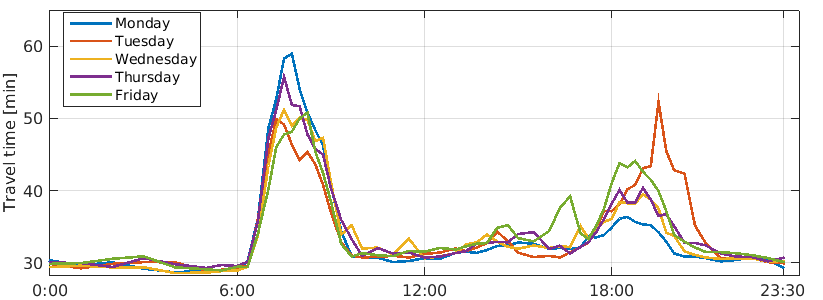}\\[4mm]
\includegraphics[width=0.95\textwidth]{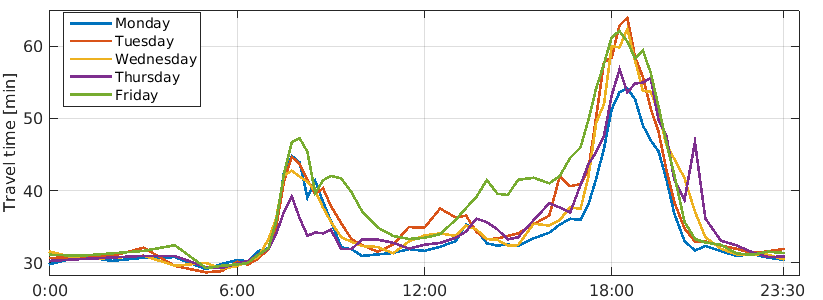} 
\caption{Travel time along Routes 11 (top) and 32 (bottom) between February 13 and 17, 2017.}\label{fig:morones-const}
\end{figure}

Figure~\ref{fig:pm10} shows graphically the data obtained of a search of PM10 at all monitoring stations (10) for the month January 2017 (TIMESTAMP\_W BETWEEN 2017-01-01 AND 2017-01-31). This search reported 6,600 observations (rows) and 3 variables (pm10, timestamp, location\_id) and it took 0.0828 seconds.

\begin{figure}[h!]
\centering
\includegraphics[width=0.95\textwidth]{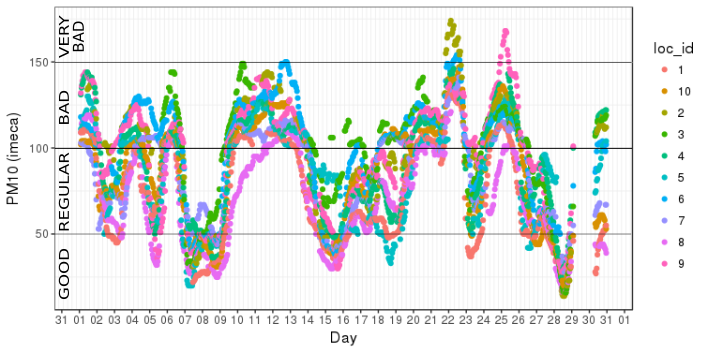}  
\caption{\label{fig:pm10} Records of PM10 of all monitoring station (10) during January 2017.}
\end{figure}

\section{Conclusions and Future Work}\label{sec:last}

This report presented how weather, traffic and air pollution data of the metropolitan area of Monterrey, Mexico, has been collected and structured into a database. R scripts were implemented to acquire data from different on-line sources and the data is paste into a structured MySQL Database.

To the best of our knowledge this is the only compendium with weather, traffic and air pollution data from of the AMM in a single database. Data visualization will be publicly available shortly on the Monterrey - Traffic, Weathers and Pollutions Database website \cite{mwtp}. Anyone that requires particular sets of data can follow the instruction found in our webpage to request it.

As future work, we would like to developed algorithm to detect multivariate outliers, as well as to develop missing-data imputation algorithms for multivariate data.

As mentioned before, the motivation of collecting the data of the MWTP Database is do to the sever pollution levels that have been reported on the AMM. Therefore, we would like to analyze, for example, the correlation between car traffic congestion and air pollution.

To complement the current database, We would like to collect information such industry air pollution, respiratory conditions, cancer, among others.

\appendix
\section*{Appendix A: Attribute tables}
Tables \ref{tab:WInfo}, \ref{tab:TInfo} and \ref{tab:PInfo} show the name of the attributes and a short description for the three main subsets within the database. Additionally, Table \ref{tab:Locations} show the attributes of the tables with location information for weather stations, traffic routes, and pollution stations. Table \ref{tab:ExtraInfW} shows the attributes information of the extra tables of weather attributes.

\begin{table}
\begin{center}
\begin{small}
\begin{tabular}{ l l }
\hline
Attribute Name  & Description (units)\\
\hline
ID\_WEATHER	  	    & weathers table identifier (primary key)\\
TIMESTAMP\_W		& Time stamp weather data (YYYY-MM-DD HH:MM:SS) (rows)\\
ID\_TIME\_ZONE		& Time zone (foreign key) \\
TEMP				& Temperature (C)\\
DEWPT				& Dew Point	(C)	\\
HUM					& Humidity (\%) \\
WSPD				& Wind Speed (km/h)	\\
WGUST				& Wind Gust	 (km/h)\\
WDIRD				& Wind Direction (degress)\\
ID\_WDIRE			& Wind Direction Description (i.e. SW, NNE, etc.) (foreign key)\\
PRESSURE			& Pressure (hPa) \\
WINDCHILL			& Wind Chill (C) \\
HEATINDEX			& Heat Index (C)	\\
PRECIPRATE			& Precipitation Rate (mm/h)\\
PRECIPTOTAL			& Accumulated daily precipitation	(mm)\\
SOLARRADIATION		& Solar Radiation (W/m$^2$)\\
UV					& UV Index (1 to 11)\\
VIS					& Visibility	(km), only available at airports\\
PRECIP				& Precipitation	(mm), only available at airports	\\
ID\_COND			& Table of Condition (i.e. Clear, Partly Cloudy, etc.)(foreign key).\\
ID\_ICON			& As Condition (foreign key).\\
FOG					&	Fog presence (0,1), only available at airports	\\
RAIN				&	Rain presence	(0,1), only available at airports	\\
SNOW				&	Snow presence (0,1), only available at airports	\\
HAIL				&	Hail presence (0,1), only available at airports	\\
THUNDER				&	Thunder presence (0,1), only available at airports	\\
TORNADO				&	Tornado presence (0,1), only available at airports\\
METAR				&	METAR raw data, only available at airports\\
ID\_LOCATIONS\_W	&  locations\_w identifier (foreign key)\\
\hline
\end{tabular}
\end{small}
\end{center}
\caption{Attributes: Weathers Table}
\label{tab:WInfo}
\end{table}

\begin{table}
\begin{center}
\begin{small}
\begin{tabular}{ l l}
\hline
Attribute Name  & Description (units)\\
\hline
ID\_TRAFFIC			& traffics table identifier (primary key)\\
TIMESTAMP\_T		& Time stamp traffic data (YYYY-MM-DD HH:MM:SS) (rows)\\
TRAVELDIST			& Travel distance (m)\\
TRAVELTIME\_STD	    & Standard travel time	(s)\\
TRAVELTIME\_CURR	& Current travel time	(s)\\
ID\_ LOCATIONS\_T	& locations\_t table identifier (foreign key)\\
\hline
\end{tabular}
\end{small}
\end{center}
\caption{Attributes: Traffic Table}
\label{tab:TInfo}
\end{table}

\begin{table}
\begin{center}
\begin{small}
\begin{tabular}{ l l}
\hline
Attribute Name  & Description (units)\\
\hline
ID\_POLLUTION	&	pollutions table identifier (primary key)\\
TIMESTAMP\_P	&	Time stamp pollution data (YYYY-MM-DD HH:MM:SS) (rows)\\
PM10	       	&	Particulate matter smaller than 10$\mu$m, PM10 (IMECA) \\
O3				&	Ozone, O3 (IMECA)\\
CO				&	Carbon Monoxide, CO (IMECA)\\
SO2				&	Sulfur Dioxide, S02 (IMECA)\\
NO3				&	Nitrogen Dioxide, NO2 (IMECA)\\
PM25			&	Particulate matter smaller than 2.5$\mu$m, PM2.5 (IMECA)\\
ID\_ LOCATIONS\_P &	locations\_p table identifier (foreign key)\\
\hline
\end{tabular}
\end{small}
\end{center}
\caption{Attributes: Pollutions Table}
\label{tab:PInfo}
\end{table}

\begin{table}
\begin{center}
\begin{small}
\begin{tabular}{ l l }
\hline
Attribute Name  & Description\\
\hline
\textbf{location\_w } & \\
ID\_LOCATIONS\_W	&	locations\_w table identifier (primary key)\\
FILE\_ID			&	Weather station name defined by user	\\
STATION\_ID			&	PWS ID defined by Weather Underground\\
AIRPOT\_CODE		&	Airport code defined by Weather Underground\\
LAT					&	Latitude of station's location.	\\
LONG				&	Longitude of station's location.	\\
DESCRIPTION			&	Short station description defined by user	\\
SOFTEARE\_TYPE		&	Software used to record data on PWS\\
SINCE				&	Date of first record per stations\\
\textbf{location\_t } & \\
ID\_LOCATIONS\_T	&	locations\_t table identifier (primary key)	\\
FILE\_ID			&	Name of route defined by the user	\\
START\_LAT			&	Latitude of start point of route	\\
START\_LONG			&	Longitude of start point of route	\\
END\_LAT			&	Latitude of end point of route	\\
END\_LONG			&	Longitude of end point of route\\
DESCRIPTION\_FROM	&	Short description of start point of route 	\\
DESCRIPTION\_TO		&	Short description of end point of route 	\\
\textbf{location\_p } & \\
ID\_LOCATIONS\_P	&	locations\_p table identifier (primary key).	\\
FILE\_ID			&	Air pollution station name defined by user	\\
LAT					&	Latitude of station's location.	\\
LONG				&	Longitude of station's location.	\\
DESCRIPTION			&	Short station description defined by the user	\\
\hline
\end{tabular}
\end{small}
\end{center}
\caption{Attributes of location\_w(location of weather stations), location\_t (traffic routes), and location\_p (location of pollution stations) tables}
\label{tab:Locations}
\end{table}

\begin{table}
\begin{center}
\begin{small}
\begin{tabular}{ l l}
\hline
Attribute Name  & Description\\
\hline
\textbf{Table: times\_zones} & \\
ID\_TIME\_ZONE			&	time\_zones table identifier (primary key)	\\
TIME\_ZONE				&	Time zone name (i.e. CST, EST etc.)	\\
DESCRIPTION				&	Description of the time zone	\\
\textbf{Table: wdires} 	&  \\
ID\_WDIRE				&	wdires table identifier (primary key)	\\
WDIRE					&	Wind direction description (i.e. SW, NNE, etc.)\\
DESCRIPTION				&	Description of wind direction type 	\\
\textbf{Table: conds} 	& \\
ID\_COND				& Conds table identifier (primery key).	\\
COND					& Name of climate condition (i.e. Clear, Partly Cloudy, etc.)	\\
DESCRIPTION				& Climate condition description \\
\textbf{Table: iconds} 	& \\
ID\_ICON				&	icons table identifier (primary key).	\\
ICON					&	Climate condition icon name	(i.e. Clear, Partly Cloudy, etc.)\\
DESCRIPTION				&	Climate condition icon description 	\\
\hline Search online for a list of available styles.
\end{tabular}
\end{small}
\end{center}
\caption{Attributes: Weather depended tables}
\label{tab:ExtraInfW}
\end{table}

\section*{Acknowledgments}
The authors gratefully acknowledge the support of the Mexican National Council for Science and Technology (CONACYT) through the C\'atedras CONACYT Program, Project 3162 \textit{"Modelaci\'on Matem\'atica, Estad\'istica y Computacional para datos complejos en un contexto de Big Data"}.

\bibliography{main}

\bibliographystyle{naturemag-doi}

\end{document}